# Analyzing the Single Event Upset Vulnerability of Binarized Neural Networks on SRAM FPGAs

Ioanna Souvatzoglou, Athanasios Papadimitriou, Aitzan Sari, Vasileios Vlagkoulis, Mihalis Psarakis

Dept. of Informatics, University of Piraeus, Greece

{ isouvatz | thanospap | aitsar | v.vlagkoulis | mpsarak }@unipi.gr

*Abstract*— **Neural Networks (NNs) are increasingly used in the last decade in several demanding applications, such as object detection and classification, autonomous driving, etc. Among different computing platforms for implementing NNs, FPGAs have multiple advantages due to design flexibility and high performance-to-watt ratio. Moreover, approximation techniques, such as quantization, have been introduced, which reduce the computational and storage requirements, thus enabling the integration of larger NNs into FPGA devices. On the other hand, FPGAs are sensitive to radiation-induced Single Event Upsets (SEUs). In this work, we perform an in-depth reliability analysis in an FPGA-based Binarized Fully Connected Neural Network (BNN) accelerator running a statistical fault injection campaign. The BNN benchmark has been produced by FINN, an open-source framework that provides an end-to-end flow from abstract level to design, making it easy to design customized FPGA NN accelerators, while it also supports various approximation techniques. The campaign includes the injection of faults in the configuration memory of a state-of-the-art Xilinx Ultrascale+ FPGA running the BNN, as well an exhaustive fault injection in the user flip flops. We have analyzed the fault injection results characterizing the SEU vulnerability of the circuit per network layer, per clock cycle, and register. In general, the results show that the BNNs are inherently resilient to soft errors, since a low portion of SEUs in the configuration memory and the flip flops, cause system crashes or misclassification errors.**

## I. Introduction

Research in Artificial Intelligence (AI) has dramatically developed in recent years, driven by theoretical advances in Machine Learning (ML) and the introduction of large-scale Deep Neural Networks (DNNs). Given that the dimensions of deeper networks push the computational and memory requirements to the limit, efficient computing platforms are needed to deal with this increasing complexity. GPUs, ASICs, and FPGAs are the most well-known hardware architectures that provide DNN acceleration by exploiting the inherent parallelism of neural networks [1]. The easy programming of commodity GPUs makes them the most well-established computing platform in the ML community. On the other hand, ASICs and FPGAs provide a higher performance/watt ratio than GPUs because they can be highly customized according to specific structural and functional requirements. Moreover, due to their high flexibility, FPGAs can be efficiently tailored to fit the irregular parallelism or the low-precision data types imposed by recently proposed approximation techniques, such as pruning and quantization. Thus, SRAM FPGAs provide the most promising computing platform for accelerating the future generation of DNNs [2], [3]. At the same time, several FPGA-based accelerators have been presented integrating software and/or hardware optimization techniques to improve performance and power consumption [4].

The approximation techniques are inspired by the inherent resiliency of NNs to the reduction of computational precision. In other words, the computational accuracy can be "sacrificed" for the sake of reducing the computational requirements, which, however, slightly affects the DNN classification accuracy. Typical approximation techniques are *quantization*, which reduces the data precision of the weights and/or neuron outputs and *weight-reduction*, which removes the redundant network parameters through structural simplification (e.g., pruning) [5]. Recent approaches take advantage of these approximation techniques to propose well-optimized FPGA-based accelerators [6],[7],[8],[9]. By eliminating the need for floating-point arithmetic, these approaches enable the integration of larger DNNs in the FPGA device, and at the same time, increase the speed and the energy efficiency of the accelerators, with low impact on the inference accuracy.

Moreover, various frameworks have been presented that support the automated mapping of DNNs to optimized FPGA accelerators [10]. These DNN-to-FPGA toolflows enable the implementation of DNN hardware accelerators without requiring FPGA design expertise and, thus, further motivate the adoption of FPGAs within the ML community. One such open-source framework, developed by Xilinx Research Labs, is FINN [7]. It provides an end-to-end design environment to automate the design exploration and the development of Binarized Neural Networks (BNNs) into Xilinx FPGAs. FINN-R [8] is an extended version of the tool that supports quantization (e.g., arbitrary data precision) and allows the user to configure more end architecture parameters. Besides the quantization & binarization, it incorporates more optimization techniques, such as activation and weight reduction and folding.

However, SRAM FPGAs are particularly vulnerable to radiation-induced Single Event Upsets (SEUs) due to their high reliance on SRAM technology. Thus, to use FPGA-based accelerators in critical applications operating in harsh environments with high radiation doses (e.g., satellites), we must analyze their SEU vulnerability and study efficient SEU mitigation approaches [11]. On the other hand, it is believed that the adoption of neural networks in onboard computers has the potential to revolutionize earth observation and remote sensing, assisting the deployment of new satellite applications [12]. Towards this end, it is vital to analyze the SEU vulnerability of FPGA-based accelerators for DNNs.

Recent approaches study the reliability of FPGA-based DNN accelerators using emulated or laser-based fault injection or/and radiation experiments [13]. In [14], the authors examined the vulnerability of RTL NN accelerators against permanent and transient faults and proposed a low-overhead fault mitigation technique. In [15], the authors conducted experiments on NNs implementations (MNIST dataset) on three different computing platforms (FPGAs, GPUs and Intel Xeon

Phis) and analyzed their reliability with respect to data precision. In [16], Libano et al. evaluated the effects of radiation-induced faults in the correct operation of two FPGA NNs (Iris Flower ANN and MNIST CNN) and proposed a selective hardware approach to the most vulnerable layers. Later, they examined the combined impact of reduced data precision and soft errors on the convolutional layers of FPGA-based NNs through fault injection [17] or neutron radiation experiments [18]. All these works were demonstrated in simple hardware designs without considering a state-of-the-art DNN-to-FPGA design flow or an FPGA-optimized DNN architecture. Recently, Gambardella et al. [19] studied the impact of permanent faults on a more sophisticated Quantized Neural Network (QNN) architecture, specifically the FINN accelerator, and proposed two different methods, a selective replication technique and a fault-aware scheduling solution, to improve the robustness of the QNN. The impact of radiation-induced SEUs on the reliability of QNN was not addressed in this work. Finally, in [20] the authors investigated the impact of Single Bit Upsets (SBUs) and Multiple Bit Upsets (MBUs) striking the memory words that store the weights and the activations of a BNN (modified version of FINN) accelerator on the classification accuracy of the network. It must be noted that this work examined the SEUs only on network parameters stored in off-chip or on-chip memories and not in the registers of the BNN or the FPGA configuration memory. Given that the memories can be protected by error correction codes and the fact that the configuration memory occupies the largest portion of the FPGA embedded memories, it is a must to study the vulnerability of the FPGA BNNs against SEUs in the configuration memory and the user registers.

In this paper, we analyze in-depth the SEU vulnerability of an FPGA BNN accelerator through an extensive fault injection campaign. A binarized version of FINN for the MNIST dataset is used as a demonstration vehicle. We perform a statistical fault injection campaign on the configuration memory of the BNN accelerator using an emulation-based fault injection technique and an exhaustive fault injection campaign for the flip flops using a simulation-based method for better observability. The main contributions of the current work are:

- The analysis is performed in a state-of-the-art BNN accelerator, such as FINN, that integrates several architectural mechanisms that it would be interesting to be studied from a performance vs. reliability point of view.
- The FINN is supported by an end-to-end DNN-to-FPGA design flow that enables the execution of a cross-layer reliability analysis in the future.
- The fault injection campaign targets both the FPGA configuration memory and the user flip-flops, while the effects of faults are analyzed per network layer, clock cycle, and register, providing many helpful insights for the SEU vulnerability of the design.
- The experiments showed that the BNN FINN accelerator has an inherent resiliency to soft errors that deserves to be further analyzed in the future.

## II. FPGA-BASED BNN ACCELERATOR

### A. Quantized Neural Networks

A Deep Neural Network (DNN) is an artificial neural network with at least one hidden layer in between. DNNs consist of three parts: input layer, hidden layers and output layer, with each layer composed of multiple neurons. In fully connected networks, the neurons of each layer are connected with every neuron of the previous layer. The connections between neurons are assigned weights, which are being multiplied with the input values to calculate the activations. Traditionally, DNNs use floating-point representation of data, but it has been proved that the high precision of floating-point numbers does not actually contribute to higher accuracy in several classification problems. Alternatively, Quantized Neural Networks (QNNs), which adopt an approximate representation of data, consume less power than the classic NNs and reduce significantly both memory and computing requirements. QNNs use bitwise operations to perform the forward and back propagation, which can be implemented and parallelized more effectively in hardware than the floating-point operations. There are different quantization techniques [21], like pruning, rounding, probabilistic, etc. In this paper, we focus on the approximation of the weights and activations, where the 32-bit data values are quantized with fewer bits. DNNs using 1-bit to represent these values are called Binarized Neural Networks (BNNs) and are studied here. BNNs are a promising solution for implementing large-scale DNN accelerators in FPGA technology since the data binarization enables the integration of deeper NNs in single FPGA devices with a slight degradation in the network accuracy.

### B. FINN

FINN is an open-source design framework from Xilinx Research Labs [7][8] that can be used to explore the design space of FPGA-based QNNs. It interfaces with various neural network training frameworks, such as Theano, Caffe, DarkNet, and Tensorflow and provides an end-to-end automated design flow that can generate customized dataflow-style architectures for different network topologies. Although there are several attempts to port QNNs on FPGA technology, most of them cannot be reused by other research teams, either due to lack of open-source code or because they are not supported by an automated DNN-to-FPGA toolflow that would enable the exploration of different DNN topologies and approximation techniques [22]. On the other hand, FINN is publicly available and actively maintained by Xilinx Research Labs. FINN provides a built-in HLS hardware library that supports a wide range of data precisions and streaming components for implementing different layer architectures, such as fully connected, convolutional, pooling, etc. Another advantage is that FINN allows the network to be fully customized, being able to determine different parameters, such as Number of Processing Elements (PEs), Number of Single Instruction Multiple Data (SIMD), folding and FIFOs size for each layer individually. For all the above reasons, we believe that FINN provides an appropriate platform to study the impact of radiation-induced SEUs on QNNs and evaluate, in the future, various SEU mitigation approaches.

### C. BNN MNIST

Our analysis relies on a pre-trained fully connected network for classifying the Modified National Institute of Standards and Technology (MNIST) dataset. The MNIST dataset consists of 28x28 grayscale pixel images of hand-written decimal digits (from 0 to 9). The network receives a 784x1 matrix as input and produces a 10-element vector as output, one float number per digit. Note that the input images are binarized, i.e., zero pixels are converted to '0' and non-zero pixels to '1', to form the binary input vector. Our BNN consists of the input layer (Layer 0), three hidden fully connected layers (Layers 1, 2 and 3), and the output layer (Tlast), as shown in Fig. 1. The accuracy of the BNN is about 93%.

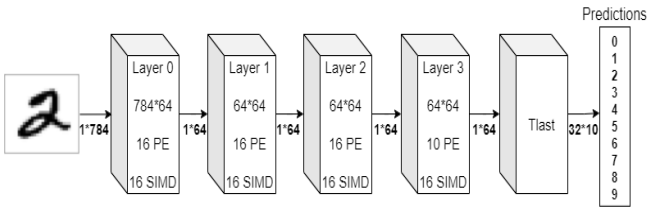

Fig. 1. BNN MNIST topology

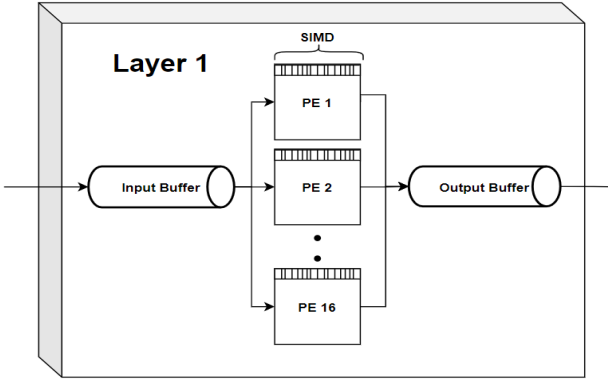

Fig. 2. Layer 1 structure

Each fully connected layer has a predefined number of Processing Elements (PEs), Single Instruction Multiple Data (SIMD) engines, and input/output buffers for data streaming. For example, Fig. 2 depicts the structure of Layer 1, which contains one input buffer, 16 Processing Elements with 16 SIMD each, and one output buffer. Each PE computes in parallel $n$ neuron products, where $n$ is the number of SIMD engines. The design parameters (synapses, PEs, SIMD units) of each layer are shown in Fig. 1.

Our BNN model has been implemented on a Xilinx UltraScale+ FPGA device. The hardware resources (LUTs, FFs, BRAMs) of the synthesized BNN accelerator are shown in Table I. To allow the injection of faults in the configuration memory – and thus the reliability analysis - per layer, the BNN design was partitioned into components, one per layer, which were placed into specific regions of the FPGA layout. The last two columns of Table I present the number of configuration memory (CRAM) bits and essential CRAM bits for every component. (Essential CRAM bits are the configuration memory bits that - when being flipped due to an SEU - have a high probability of producing incorrect output. This information was extracted from the .ebd (Essential Bits Description) file generated by Vivado implementation tools.)

## III. FAULT INJECTION CAMPAIGN

In this section, we present the strategy of our fault injection campaigns including the adopted fault model and the Fault Injection framework.

TABLE I.   HARDWARE RESOURCES OF THE BNN ACCELERATOR

|         | LUTs | FFs  | BRAMs | CRAM Bits | Essential CRAM Bits |
|---------|------|------|-------|-----------|---------------------|
| Layer 0 | 1743 | 1648 | 8     | 1601088   | 573329              |
| Layer 1 | 1714 | 1179 | -     | 2083200   | 383069              |
| Layer 2 | 1705 | 1177 | -     | 2862912   | 407612              |
| Layer 3 | 988  | 1164 | -     | 1505856   | 356114              |
| Tlast   | 661  | 1294 | -     | 2196288   | 220192              |
| **Total** | **6811** | **6462** | **8** | **10249344** | **1940316**     |

### A. Fault Model

The fault model adopted in our experiments is Single Event Upset (SEUs) in the configuration memory and the user flip-flops (FFs). Given that the configuration memory occupies a vast portion of the memories embedded in the FPGA devices, it is evident that it has a higher impact on the SEU vulnerability of the BNN accelerator compared to the other device memories. Thus, the SEUs in the configuration memory are the dominant fault model in our reliability analysis. We inject upsets only in the essential memory configuration bits to save time since the non-essential bits are not relevant to the operation of the accelerator. Note that less than 20% of the total configuration bits of the core are essential, as shown in Table I. Moreover, we consider the worst-case scenario since the fault is injected in the CRAM before the accelerator starts processing the input image and remains active for the entire duration of the current classification run (Note: there is no overlapping between the consecutive classification runs). This way, we reduce the target fault space since we do not need to examine the activation of an upset in a specific configuration bit for several timing windows but only once. On the other hand, it is a realistic scenario since a single classification run lasts a few hundreds of clock cycles, which is much shorter than the average error correction latency of any scrubbing mechanism that could be used to protect the configuration memory. Due to the large number of sensitive configuration bits, we perform statistical fault injection as described in the experimental section. Concerning the SEUs in the FFs, we adopt an exhaustive fault model since we inject several faults for every FF of the accelerator, i.e., one SEU in each clock cycle of a single classification run.

### B. Fault Injection Framework

#### 1) Configuration Memory Fault Injection

The setup for the Configuration Memory fault injection campaign consists of a PC and an Ultra96 ZU3EG Ultrascale+ Xilinx FPGA. The NN is implemented inside the Programmable Logic (PL) of the FPGA, while the Processing System (PS) is responsible to provide an interface with the NN. Communication between the Ultra96 board and the PC is achieved by means of an ethernet connection. To perform the emulation-based fault injection in the CRAM we use the FPGA REliability Evaluation through JTAG (FREtZ) framework. FREtZ is an open-source tool [23],[24] integrating a python front-end application that can use Vivado TCL commands and communicate with the Xilinx FPGAs through the JTAG port. Initially, FREtZ can extract from the design all the sensitive unmasked bits of the CRAM, including their frame addresses and bit locations. Furthermore, it is capable of injecting faults into the CRAM according to our fault model. The fault injection process in the CRAM is as follows. The FPGA device is programmed and the test image is downloaded through the Ethernet connection. Next, the fault is injected and the BNN is triggered to run the classification process. At the end of the run, the classification results are recorded in the PC. A timeout condition is used to detect the cases that the BNN has been crashed. The host application compares the results with the golden signature, and if it detects an error or a crash condition, it resets (reconfigures) the device to avoid the malfunction of the subsequent tests due to the possible occurrence of a SEFI condition. Otherwise, the fault is corrected and a new fault injection cycle starts. The CRAM Fault injection campaign setup is presented in Fig. 3.

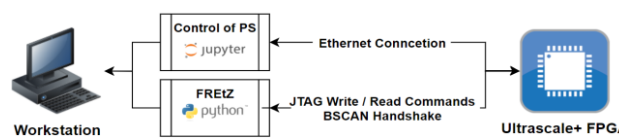

Fig. 3. Physical Setup for the CRAM Fault Injection Campaign

In this analysis, we are interested to characterize the SEU vulnerability of the complete BNN, but also of each network layer individually. This analysis would help us in the future to form an efficient, customized fault mitigation solution. To achieve this, the first step is to locate each layer in a separate partition using placement constraints. This way, we are able to correlate the configuration memory frames to be injected with faults with the hardware logic belonging to specific BNN layers. Then, we use FREtZ in order to retrieve the fault list for the CRAM. The framework identifies all the essential non-masked, configuration bits of every BNN layer (masked bits, i.e. CRAM bits associated with the storage elements of the design, such as FFs, SRLs, LUTRAMs, and BRAMs, are excluded from the fault injection process). Our BNN accelerator contains 3443 CRAM frames with 10249344 CRAM bits, from which the 1940316 are the sensitive unmasked CRAM bits that constitute the fault list for an exhaustive fault injection campaign. For reference, we provide the implementation details for this design in Table I.

Based on [25], we perform a statistical analysis of a confidence interval of 99%, while we perform fault injection to achieve a minimum margin of error of 0.3%. These values correspond to the injection of approximately 10% of the complete fault list (194031 faults).

*2) Flip Flop Fault Injection*

The fault injection campaign on FFs was conducted by means of simulation commands in the Questasim Simulator. We inject the SEUs in the FFs by using force commands on their inputs. In Table I, the FF resources of our BNN, 6462 in total, and of each layer are presented. The complete fault list includes 6462*235 faults (235 is the number of clock cycles in a single classification run).

## IV. EXPERIMENTAL RESULTS

In this section we analyze the results of the fault injection campaigns. The faults are categorized similarly to [16],[17], as a) *Critical Faults*, which lead to misclassifications, b) *Tolerable Faults*, where we observe faults in the output, but the classification of the NN remains correct, c) *Crashes*, where the system fails to execute the classification test and crashes. These categories are analyzed per layer for the CRAM and per layer, per cycle and per register for the FFs.

*A. Configuration Memory*

Using the setup described in Section III-B.1, we first performed a statistical fault injection campaign on the configuration memory of our BNN with a confidence level of 99% and a margin of error (MoE) 0.3%. The results, categorized as tolerable errors, crashes and critical errors, are shown in Table II. Analyzing the results of this first campaign, we observe the following: i) Most of the injected faults (~93.6%) do not lead to the propagation of errors to the BNN outputs. ii) For the remaining ~6.4%, the majority of faults lead to tolerable errors (~5%), while there occurred ~1.2% of crashes and ~0.2% of critical errors. iii) Analyzing the results per layer, we conclude that they are approximately the same for all layers except the final Tlast. A more detailed look shows that Layer1 and Layer2 produce more tolerable errors and slightly more crashes, while Layer2 and Layer3 produce more critical errors. At this point, we should mention that even though the design produces a 32-bit output for each class, only the 6 LSB bits contain useful, non-zero, information. Therefore, all the errors affecting the 26 Most Significant Bits (MSB) of the ten outputs are considered as redundant and have been removed from our analysis.

In order to validate the statistical analysis, we performed two additional fault injection campaigns keeping the confidence level constant at 99% and increasing the MoE at 1% and 5%; the results are included in Table II. Comparing the three campaigns, we notice that the results in total are similar in all cases. On the other hand, increasing the MoE to 1% the results per layer are similar too, but above it (MoE 5%) we see notable differences in some cases. This is evidence that we can trust the statistical results from a campaign with 1% MoE, which needs the injection of approximately 16k faults, instead of 194k faults required for 0.3% MoE (or the exhaustive campaign of 1.9M faults), but not from a campaign with 5% MoE or more. This observation will be helpful for our future work, including several fault injection experiments with different NNs and approximation techniques.

TABLE II. CRAM FAULT INJECTION CAMPAIGN RESULTS

| CRAM | Tolerable faults | | | Crashes | | | Critical faults | | |
|---|---|---|---|---|---|---|---|---|---|
| | MoE | | | | | | | | |
| | 0.3% | 1% | 5% | 0.3% | 1% | 5% | 0.3% | 1% | 5% |
| Total | 4.98 | 4.80 | 4.97 | 1.06 | 1.18 | 1.20 | 0.23 | 0.23 | 0.30 |
| Layer 0 | 3.42 | 3.76 | 4.86 | 1.18 | 1.29 | 1.62 | 0.19 | 0.31 | 0.54 |
| Layer 1 | 6.10 | 5.81 | 6.77 | 1.48 | 1.42 | 1.69 | 0.21 | 0.25 | 0.00 |
| Layer 2 | 5.94 | 6.60 | 5.71 | 1.29 | 1.27 | 0.00 | 0.18 | 0.18 | 0.00 |
| Layer 3 | 3.45 | 4.17 | 3.22 | 1.23 | 1.19 | 3.22 | 0.26 | 0.32 | 1.07 |
| Tlast | 1.57 | 1.80 | 1.75 | 0.56 | 0.41 | 0.00 | 0.19 | 0.20 | 0.00 |

TABLE III. FF FAULT INJECTION CAMPAIGN RESULTS (Exh means Exhaustive)

| FFs | Cycles duration | | | Tolerable faults | | | | Crashes | | | | Critical faults | | | |
|---|---|---|---|---|---|---|---|---|---|---|---|---|---|---|---|
| | # | Start | End | Exh | Statistical (MoE) | | | Exh | Statistical (MoE) | | | Exh | Statistical (MoE) | | |
| | | | | | 0.3% | 1% | 5% | | 0.3% | 1% | 5% | | 0.3% | 1% | 5% |
| Total | 235 | 1 | 235 | 2.69 | 2.72 | 2.84 | 1.51 | 2.25 | 2.19 | 2.12 | 1.66 | 0.12 | 0.12 | 0.10 | 0.00 |
| L0 | 198 | 1 | 198 | 8.54 | 8.52 | 8.43 | 4.97 | 2.38 | 2.34 | 2.32 | 1.86 | 0.45 | 0.42 | 0.40 | 0.00 |
| L1 | 164 | 51 | 214 | 1.03 | 1.15 | 1.53 | 0.00 | 2.95 | 2.75 | 2.12 | 0.80 | 0.00 | 0.00 | 0.00 | 0.00 |
| L2 | 29 | 202 | 230 | 1.70 | 1.67 | 2.08 | 0.75 | 2.94 | 2.86 | 2.88 | 0.75 | 0.02 | 0.02 | 0.03 | 0.00 |
| L3 | 16 | 218 | 233 | 0.13 | 0.10 | 0.06 | 0.86 | 3.06 | 3.00 | 3.36 | 5.17 | 0.03 | 0.03 | 0.00 | 0.00 |
| Tlast | 3 | 233 | 235 | 0.01 | 0.01 | 0.00 | 0.00 | 0.15 | 0.15 | 0.09 | 0.00 | 0.00 | 0.00 | 0.00 | 0.00 |

## B. Flip Flops

The analysis for this campaign is conducted driven by three factors: a) per layer, where we present the results of the three error categories per layer, b) per cycle, where we observe the impact of faults in different phases of the inference process, c) per register, where we examine possible vulnerabilities of individual FFs. Table III presents the results of an exhaustive fault injection campaign, as well as three extra statistical campaigns of 0.3%, 1% and 5% MoE. The first three columns of the table provide the duration of a single classification process in clock cycles, as well as the duration and the active period of each layer. Concerning the exhaustive campaign for the entire design, we can see that ~95% of the injected faults did not lead to an error at the circuit's output. Furthermore, the majority of errors belong to the categories of either tolerable (~2.7%) or crashes (~2.2%), while the critical errors are limited to approximately 0.12%. By considering the fault injections per layer, we can see that the highest portion of tolerable and critical errors are by far in Layer0. On the other hand, Layers3 and Tlast produce considerably less tolerable errors than the previous layers. Therefore, both tolerable and critical errors show that the most sensitive/vulnerable layer to the injection of faults is Layer0. This result could be explained by the fact that Layer0 is larger than all the other layers, in terms of both the number of FFs (see Table I) and duration (see Table III) and its results spread out in subsequent layers. Once again, we notice that the statistical injection campaigns considering 0.3% and 1% MoE reveal similar percentages for all error categories. On the other hand, when the MoE increases to 5% we can clearly see that the results are significantly different in some cases.

To gain a better insight into the relation of the injected faults per layer and time, we have separated the clock cycles of a single classification run into distinct phases. The first phase starts at the beginning of the BNN inference process along with Layer0. Then, we define a new phase each time a new layer is starting its operation until the end of the classification. Table IV presents all the relevant phases for this particular NN, including the starting and ending clock cycles for each of them. The last column of Table V shows which are the active layers during each phase.

In Fig. 4, we further categorize the faults reported in Table III according to the phase and layer they belong to. The phases are separated in bars while the layers are illustrated with different colors. The vertical axes show the percentage of faults injected in each phase, which led to either tolerable, critical or crash errors. Concerning crashes, this analysis shows that they may occur, even if a fault is injected in the FF of a layer that has been not started yet. For instance, faults injected in Layer3 during the first phase may lead to crashes. This could be explained when faults are injected in the control logic of the layer affecting its state machine, and thus, its operation in the next phases. Additionally, the vast majority of critical errors in FFs of Layer0 are occurring during the first phase showing that Layer1 FFs are sensitive to critical faults in the first fifty clock cycles. In general, critical errors for all layers were mainly the result of faults injected in the starting phase of each layer.

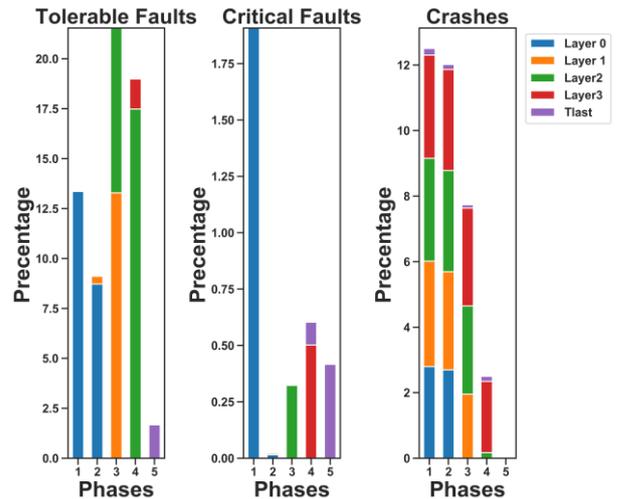

Fig. 4. Flip Flops Results per Phase

Fig. 5 presents the histograms of the faults per FF for the three categories. The colors are used to show the layer that the FFs belong to. Vertical axis denotes the number of faults that, when injected in the specific FF (in different clock cycles), caused the corresponding type of error. The histograms can reveal the criticality of each FF, i.e., its impact on the SEU vulnerability of the design. First, the histograms confirm the initial observation that the source of most critical and tolerable errors is Layer0. Second, they can provide an insight on the most critical FFs, which can be carefully inspected after such an analysis and protected by a well-optimized SEU mitigation approach. For instance, the inspection of the critical faults histogram showed that there is a 32-bit register in Layer0 which causes the majority of critical faults in this layer. By parsing the netlist, we found that this register is used to construct the memory addresses that Layer0 accesses to read the weights of its input synapses. Therefore, any upset in this register forces the NN to read wrong weights, heavily affecting the network accuracy. Given that such registers can be easily protected with negligible overhead, if we exclude the critical faults caused due to this register, the portion of total critical faults is reduced from 0.45% to 0.04%.

TABLE IV.  ACTIVE LAYERS PER PHASE

| Active Layers per Phase | | | |
|---|---|---|---|
| | Start | End | Active Layers |
| Phase 1 | 1 | 50 | Layer 0 |
| Phase 2 | 51 | 201 | Layer 0, Layer 1 |
| Phase 3 | 202 | 217 | Layer 1, Layer 2 |
| Phase 4 | 218 | 232 | Layer 2, Layer 3 |
| Phase 5 | 233 | 235 | Tlast |

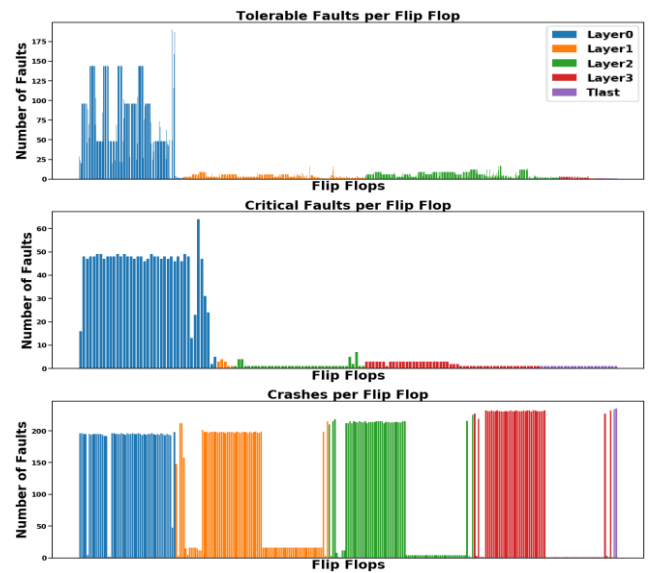

Fig. 5. No. of Faults per Flip Flop

*C. Comparison with Related Works*

In [13], Benevenuti et al. evaluated the reliability of an ANN trained for classification on the Iris flowers dataset, by performing radiation and fault injection tests. The fault injection method showed that 2.5% of the configuration bits of the accelerator are critical (i.e., bits causing functional failure). In our case, 6.27% of the faults in the essential configuration bits caused a functional error (e.g., critical error, tolerable error, or crash). But, if we project the number of errors to the total configuration bits of the BNN accelerator, the overall percentage of critical bits is 1.2%, which is less enough than those reported in [13]. On the other hand, the ratio between tolerable and critical faults (in our case, if we count the crashes as critical faults, we have 23 % critical and 77% tolerable faults) is very close to [13].

In [16], Libano et al. calculated the AVF of the hidden layers and the output layer of the ANN Iris Flower using fault injection. They reported that hidden layers produce 7% of tolerable errors and 1% of critical errors, while the output layer produces 4% and 0.5%, respectively. In our case, if we consider Layer0 to Layer3 as hidden layers and Tlast as an output layer, the AVF factors are similar: 3.41% tolerable and 0.54% critical errors plus crashes in hidden layers, and 1.57% tolerable and 0.75% critical errors plus crashed in the output layer. Note that if we project these numbers to the total configuration bits of the BNN area (and not the essential ones) the AVF parameters are considerably lower. Although our FINN BNN is much more complex than the simple ANN Iris Flower, it is apparent that its SEU vulnerability is significantly lower, fact that will be further investigated in the future.

Finally, in [17], Libano et al. evaluated the reliability of a Quantized CNN trained for classification on MNIST dataset. The quantization technique was applied in the feature extraction and not in the classification part of the network. Since our BNN is a fully connected NN, we focus on comparing it with the classification part; in [17], they mention 1.26% critical bits for the classification part of the QNN, which is very close to our results. However, our BNN accelerator produces more tolerable errors (among the faults causing functional errors, 80% are tolerable and 20% are critical) than the QNN (57% tolerable faults and 43% of critical faults). This is another indication of the high resiliency of the BNNs.

## V. CONCLUSION

In this paper, we completed a detailed SEU vulnerability analysis of a Binarized Neural Network produced from a state-of-the-art framework, FINN. By performing two major fault injection campaigns, one statistical by injecting faults on the CRAM of an Ultrascale+ FPGA, and one exhaustive simulation-based, we provided a detailed reliability analysis driven by multiple factors on Binarized Neural Networks. Our analysis showed that the BNN FINN has an inherent resiliency to soft errors and provided several useful insights that could guide the design of well-optimized SEU mitigation approaches. In our future work, we will further investigate the impact of various BNN topologies and architectural parameters (quantization, folding, etc.) on the SEU vulnerability of the design. Our goal is to examine possible patterns between different topologies and architectures and to provide efficient fault mitigation techniques for FPGA BNN accelerators.

ACKNOWLEDGMENT

This research has been partially co-financed by the European Union and Greek national funds through the Operational Program Competitiveness, Entrepreneurship and Innovation, under the call RESEARCH – CREATE – INNOVATE (project code: T1EDK-04298) and by the European Union's Horizon 2020 research and innovation programme under the Marie Sklodowska-Curie grant agreement No 895937.